# Reliability estimates for three factor score predictors

André Beauducel[1, a *]**,** Christopher Harms[1, b] and Norbert Hilger[1, c]

[1]University of Bonn, Institute of Psychology, Kaiser-Karl-Ring 9, 53111 Bonn, Germany

[a]beauducel@uni-bonn.de, [b]christopher.harms@uni-bonn.de, [c]nhilger@uni-bonn.de

February 23[th], 2016

**Abstract.** Estimates for the reliability of Thurstone's regression factor score predictor, Bartlett's factor score predictor, and McDonald's factor score predictor were proposed. As in Kuder-Richardson's formula, the reliability estimates are based on a hypothetical set of equivalent items. The reliability estimates were compared by means of simulation studies. Overall, the reliability estimates were largest for the regression score predictor, so that the reliability estimates for Bartlett's and McDonald's factor score predictor should be compared with the reliability of the regression score predictor, whenever Bartlett's or McDonald's factor score predictor are to be computed. An R-script and an SPSS-script for the computation of the respective reliability estimates is presented.

**Introduction**

Factor score predictors may be computed whenever individual scores on the factors are of interest. If, for example, decisions are made on the individual level (e.g., in personnel selection) an individual score is needed. When factor score predictors are used in order to compute individual scores, it would be helpful to know whether they are valid and reliable. The coefficient of determinacy, i.e., the correlation of the factor score predictor with the factor [1] has been related to the validity of factor score predictors [2]. However, the reliability of factor score predictors has rarely been investigated. It should be noted that the reliability of factor score predictors should not be confounded with the reliability of the factors. An index for the reliability of a factor has been proposed [3,4] and this index has been found to represent the proportion of variance due to all common factors [5], and a specific form of this index has been proposed to represent the reliability of the general factor in hierarchical models. Of course, the reliability of the factors might be of interest whenever factor models are estimated. However, it is impossible to compute the individual scores on the factors because the number of common and unique factors exceeds the number of observed variables [6]. Therefore, factor score predictors have to be computed whenever individual scores representing the factors are needed. Accordingly, the reliability of the factor score predictors should be estimated.

Moreover, specific reliability estimates of factor score predictors may depend on the factor score predictors that are considered. For example, a reliability estimate for Harman's ideal variable factor score predictor [7] have been proposed [8]. However, Thurstone's regression factor score predictor [9], Bartlett's factor score predictor [10], and McDonald's correlation-preserving factor score predictor [11] are probably more often used than Harman's ideal variable factor score predictor. Therefore, the present paper aims at proposing reliability estimates for Thurstone's regression factor score predictor, Bartlett's factor score predictor, and McDonald's correlation-preserving factor score predictor.

Moreover, the effect of the size of loadings, the number of variables, the inter-correlation of the factors, and sampling error on the reliability estimates for the three factor score predictors will be investigated by means of a simulation study. It is, however, possible that the factor model does not perfectly hold in a given sample, which is typically referred to as model error [12,13]. Accordingly, the effect of model error on the reliability estimates of the factor score predictors is also investigated by means of a simulation study. Finally, an R script is presented that allows for the computation of the reliability estimates for the factor score predictors starting from the loading pattern, the factor inter-correlations, and the item covariances.

**Definitions**

In the population, the common factor model can be defined as

$$\mathbf{x} = \mathbf{\Lambda}\mathbf{f} + \mathbf{e} \tag{1}$$

where **x** is the random vector of observations or items of order $p$. Thus, there are $p$ observed variables and **f** is the random vector of common factor scores of order $q$, **e** is the random error vector or unique vector of order $p$, and **Λ** is the factor pattern matrix of order $p$ by $q$. The factors **f**, and the unique or error vectors **e** are assumed to have an expectation zero ($\varepsilon[\mathbf{x}] = 0$, $\varepsilon[\mathbf{f}] = 0$, $\varepsilon[\mathbf{e}] =$ 0). The covariance between the factors and the error scores is assumed to be zero (Cov[**f**, **e**] = ε[**fe**′] = 0). The covariance matrix of observed variables **Σ** can be decomposed into

$$\Sigma = \Lambda \Phi \Lambda' + \Psi^2, \tag{2}$$

where **Φ** represents the $q$ by $q$ factor correlation matrix and **Ψ²** is a $p$ by $p$ diagonal matrix representing the expected covariance of the error scores **e** (Cov[**e**,**e**] = ε[**ee**′] = **Ψ²**). It is assumed that the diagonal of **Ψ²** contains only positive values and that the expectation of the non-diagonal elements is zero.

**Reliability of factor score predictors**

The starting point is Kuder and Richardson's consideration that a sum of items might be correlated with a sum of hypothetical equivalent items in order to estimate the reliability of the item sum [14]. In the following, $\mathbf{x}_1$ is an empirical set of $p$ items and $\mathbf{x}_2$ is a hypothetical set of equivalent items. According Kuder and Richardson, equivalence means that the items in the hypothetical item set are interchangeable with the items in the empirical item set. Thus, the members of each item pair (comprising an empirical item and a hypothetical item) have the same difficulty and are correlated to the extent of their respective reliabilities. This implies that the inter-item correlations within each set of items need not to be equal, i.e., that the items within each set need not to be parallel items. It is, nevertheless, instructive to present the correlation between the unit-weighted scales resulting for two equivalent sets of parallel items with unit variance in matrix form. The condition of unit item variances and parallel items implies $\Sigma_1 = \Sigma_2$ and $\varepsilon[\mathbf{x}_1 \mathbf{x}_2'] = \rho \mathbf{11}'$, where **1** is a $p \times 1$ unit-vector and $\rho$ is the inter-correlation of the items. Under these assumptions the correlation of the item sum of $\mathbf{x}_1$ with the sum of hypothetical parallel items $\mathbf{x}_2$ can be written as

$$\begin{aligned}\mathrm{Cor}[\mathbf{1}'\mathbf{x}_1, \mathbf{1}'\mathbf{x}_2] &= \mathrm{diag}(\varepsilon[\mathbf{1}'\Sigma_1\mathbf{1}])^{-1/2} \varepsilon[\mathbf{1}'\mathbf{x}_1\mathbf{x}_2'\mathbf{1}] \mathrm{diag}(\varepsilon[\mathbf{1}'\Sigma_2\mathbf{1}])^{-1/2} \\ &= \varepsilon[\mathbf{1}'\mathbf{x}_1\mathbf{x}_2'\mathbf{1}] \mathrm{diag}(\varepsilon[\mathbf{1}'\Sigma_1\mathbf{1}])^{-1}. \end{aligned} \tag{3}$$

Parallel items imply $\varepsilon[\mathbf{1}'\mathbf{x}_1\mathbf{x}_2'\mathbf{1}] = p^2 \rho$ and $\varepsilon[\mathbf{1}'\Sigma_1\mathbf{1}] = p + p(p-1)\rho.$ Equation (3) can therefore be written as

$$\mathrm{Cor}[\mathbf{1}'\mathbf{x}_1, \mathbf{1}'\mathbf{x}_2] = \frac{p\rho}{1+(p-1)\rho}. \tag{4}$$

Thus, the transformation yields Kuder-Richardson formula 17, which corresponds to the Spearman-Brown prophecy formula and to Cronbach's alpha for parallel items. In the next step, the idea of using a hypothetical set of equivalent items will be used in order to propose reliability estimates for factor score predictors. Therefore, Equation 3 will be modified in that the unit vectors are replaced by the weights that are needed for the computation of factor score predictors. Thus, the following section does not refer to parallel items, but to two sets of equivalent items.

Let $\mathbf{B}_1$ be a matrix of weights of the empirical item set $\mathbf{x}_1$. The respective factor score predictor is

$$\hat{\mathbf{f}}_1 = \mathbf{B}_1' \mathbf{x}_1. \tag{5}$$

Assuming a hypothetical equivalent item set yields $\hat{\mathbf{f}}_2 = \mathbf{B}_2' \mathbf{x}_2$. Equivalent items imply $\mathbf{B}_1 = \mathbf{B}_2$ and $\Sigma_1 = \Sigma_2$. On this basis, $\text{diag}(\text{Cor}[\hat{\mathbf{f}}_1, \hat{\mathbf{f}}_2])$ the correlation of the corresponding factor score predictors can be regarded as an estimate of the reliability of factor score predictor. The reliability estimate can be written as

$$\begin{aligned}\mathbf{R}_{ttf} &= \text{diag}(\text{diag}(\hat{\mathbf{f}}_1 \hat{\mathbf{f}}_1')^{-1/2} \hat{\mathbf{f}}_1 \hat{\mathbf{f}}_2' \text{diag}(\hat{\mathbf{f}}_2 \hat{\mathbf{f}}_2')^{-1/2}) \\ &= \text{diag}(\text{diag}(\mathbf{B}_1' \Sigma_1 \mathbf{B}_1)^{-1/2} \mathbf{B}_1' \mathbf{x}_1 \mathbf{x}_2' \mathbf{B}_2 \, \text{diag}(\mathbf{B}_2' \Sigma_2 \mathbf{B}_2)^{-1/2}) \\ &= \text{diag}(\text{diag}(\mathbf{B}_1' \Sigma_1 \mathbf{B}_1)^{-1/2} \mathbf{B}_1' \mathbf{x}_1 \mathbf{x}_2' \mathbf{B}_1 \, \text{diag}(\mathbf{B}_1' \Sigma_2 \mathbf{B}_1)^{-1/2}).\end{aligned} \tag{6}$$

It follows from Equation 6 that $\mathbf{R}_{ttf} = \mathbf{0}$ even for $\mathbf{B}_1 = \mathbf{B}_2$ and $\Sigma_1 = \Sigma_2$ if $\mathbf{x}_1 \mathbf{x}_2' = \mathbf{0}$ and that $\mathbf{R}_{ttf} = \mathbf{I}$ for $\mathbf{B}_1 = \mathbf{B}_2$ and $\mathbf{x}_1 \mathbf{x}_2' = \Sigma_1 = \Sigma_2$. Of course, a perfect reliability of the observed variables implies a perfect reliability of the factor score predictors. In the following, it will not be assumed that the items have a unit-variance and that the items are parallel, because factor analysis is very unlikely to be applied to a set of parallel items. The assumption of an equivalent set of items, however, implies that the same factor model will be found in both item sets. Although it might be questioned that a factor model can be exactly reproduced, the stability of the weights is also assumed in the Kuder-Richardson formula when unit-weighted scales are considered, so that Equation 6 corresponds to this perspective. Another general assumption is that the correlation between the equivalent items is only due to the common factors. The unique or error variance is not regarded as a cause for the correlation between the equivalent item sets ($\varepsilon[\mathbf{e}_1 \mathbf{e}_2'] = \mathbf{0}$). As Cronbach has noted, $\alpha$, as well as any other reliability coefficient based on equivalent items treats the specific content of an item as error [15]. Although other perspectives are possible, however, we follow Cronbach in treating the specific item content as error.

When the weights for different factor score predictors are entered into Equation 6, the resulting equations will represent the reliability of the respective factor score predictor. For Thurstone's regression factor score predictor the weights are $\mathbf{B}_r = \Sigma^{-1} \Lambda \Phi$. Entering these weights into Equation 6 and adding subscripts indicating the empirical and the hypothetical item sets yields

$$\begin{aligned}\mathbf{R}_{ttr} &= \text{Cor}(\hat{\mathbf{f}}_{1r}, \hat{\mathbf{f}}_{2r}) \\ &= \text{diag}(\Phi_1 \Lambda_1' \Sigma_1^{-1} \Lambda_1 \Phi_1)^{-1/2} \text{diag}(\Phi_1 \Lambda_1' \Sigma_1^{-1} \mathbf{x}_1 \mathbf{x}_2' \Sigma_2^{-1} \Lambda_2 \Phi_2) \text{diag}(\Phi_2 \Lambda_2' \Sigma_2^{-1} \Lambda_2 \Phi_2)^{-1/2}\end{aligned}. \tag{7}$$

Inserting $\mathbf{x}_1 = \Lambda_1 \mathbf{f}_1 + \Psi_1 \mathbf{e}_1$ and $\mathbf{x}_2 = \Lambda_2 \mathbf{f}_2 + \Psi_2 \mathbf{e}_2$ into Equation 7 and some transformation yields

$$\begin{aligned}\mathbf{R}_{ttr} &= \text{diag}(\Phi_1 \Lambda_1' \Sigma_1^{-1} \Lambda_1 \Phi_1)^{-1/2} \text{diag}(\Phi_1 \Lambda_1' \Sigma_1^{-1} (\Lambda_1 \mathbf{f}_1 \mathbf{f}_2' \Lambda_2' + \Lambda_1 \mathbf{f}_1 \mathbf{e}_2' \Psi_2 \\ &\quad + \Psi_1 \mathbf{e}_1 \mathbf{f}_2' \Lambda_2' + \Psi_1 \mathbf{e}_1 \mathbf{e}_2' \Psi_2) \Sigma_2^{-1} \Lambda_2 \Phi_2) \text{diag}(\Phi_2 \Lambda_2' \Sigma_2^{-1} \Lambda_2 \Phi_2)^{-1/2}.\end{aligned} \tag{8}$$

It is assumed that the same factors were measured ($\mathbf{f}_1 = \mathbf{f}_2$) and it is, moreover, assumed that the same factor model holds in the population ($\Lambda_1 = \Lambda_2, \Phi_1 = \Phi_2, \Psi_1 = \Psi_2, \Sigma_1 = \Sigma_2$). This also implies $\mathbf{f}_1 \mathbf{e}_2' = \mathbf{0}$ and $\mathbf{f}_2 \mathbf{e}_1' = \mathbf{0}$. When these conditions hold and when there is no reliable unique or error variance, i.e., when there is a zero covariance of the error scores across measurement occasions ($\varepsilon[\mathbf{e}_1\mathbf{e}_2'] = \mathbf{0}$), Equation 8 can be transformed into

$$\mathbf{R}_{ttr} = \operatorname{diag}(\mathbf{\Phi}_1\mathbf{\Lambda}_1'\mathbf{\Sigma}_1^{-1}\mathbf{\Lambda}_1\mathbf{\Phi}_1)^{-1/2}\operatorname{diag}(\mathbf{\Phi}_1\mathbf{\Lambda}_1'\mathbf{\Sigma}_1^{-1}\mathbf{\Lambda}_1\mathbf{\Phi}_1\mathbf{\Lambda}_1'\mathbf{\Sigma}_1^{-1}\mathbf{\Lambda}_1\mathbf{\Phi}_1)\operatorname{diag}(\mathbf{\Phi}_1\mathbf{\Lambda}_1'\mathbf{\Sigma}_1^{-1}\mathbf{\Lambda}_1\mathbf{\Phi}_1)^{-1/2}. \tag{9}$$

Entering $\mathbf{B}_b = \mathbf{\Psi}^{-2}\mathbf{\Lambda}(\mathbf{\Lambda}'\mathbf{\Psi}^{-2}\mathbf{\Lambda})^{-1}$ for the Bartlett's factor score predictor into Equation 6 and introducing the subscripts yields

$$\begin{aligned}\mathbf{R}_{ttb} &= \operatorname{Cor}(\hat{\mathbf{f}}_{1b}, \hat{\mathbf{f}}_{2b}) \\ &= \operatorname{diag}((\mathbf{\Lambda}_1'\mathbf{\Psi}_1^{-2}\mathbf{\Lambda}_1)^{-1}\mathbf{\Lambda}_1'\mathbf{\Psi}_1^{-2}\mathbf{\Sigma}_1\mathbf{\Psi}_1^{-2}\mathbf{\Lambda}_1(\mathbf{\Lambda}_1'\mathbf{\Psi}_1^{-2}\mathbf{\Lambda}_1)^{-1})^{-1/2} \\ &\quad \operatorname{diag}((\mathbf{\Lambda}_1'\mathbf{\Psi}_1^{-2}\mathbf{\Lambda}_1)^{-1}\mathbf{\Lambda}_1'\mathbf{\Psi}_1^{-2}\mathbf{x}_1\mathbf{x}_2'\mathbf{\Psi}_2^{-2}\mathbf{\Lambda}_2(\mathbf{\Lambda}_2'\mathbf{\Psi}_2^{-2}\mathbf{\Lambda}_2)^{-1}) \\ &\quad \operatorname{diag}((\mathbf{\Lambda}_2'\mathbf{\Psi}_2^{-2}\mathbf{\Lambda}_2)^{-1}\mathbf{\Lambda}_2'\mathbf{\Psi}_2^{-2}\mathbf{\Sigma}_2\mathbf{\Psi}_2^{-2}\mathbf{\Lambda}_2(\mathbf{\Lambda}_2'\mathbf{\Psi}_2^{-2}\mathbf{\Lambda}_2)^{-1})^{-1/2}\end{aligned} \tag{10}$$

According to $\mathbf{\Lambda}_1 = \mathbf{\Lambda}_2, \mathbf{\Phi}_1 = \mathbf{\Phi}_2, \mathbf{\Psi}_1 = \mathbf{\Psi}_2, \mathbf{\Sigma}_1 = \mathbf{\Sigma}_2, \mathbf{f}_1\mathbf{e}_2 = \mathbf{0}, \mathbf{f}_2\mathbf{e}_1 = \mathbf{0}$, and $\mathbf{e}_1\mathbf{e}_2' = \mathbf{0}$ Equation 10 can be transformed into

$$\begin{aligned}\mathbf{R}_{ttb} &= \operatorname{diag}((\mathbf{\Lambda}_1'\mathbf{\Psi}_1^{-2}\mathbf{\Lambda}_1)^{-1}\mathbf{\Lambda}_1'\mathbf{\Psi}_1^{-2}\mathbf{\Sigma}_1\mathbf{\Psi}_1^{-2}\mathbf{\Lambda}_1(\mathbf{\Lambda}_1'\mathbf{\Psi}_1^{-2}\mathbf{\Lambda}_1)^{-1})^{-1/2} \\ &\quad \operatorname{diag}(\mathbf{\Phi}_1)\operatorname{diag}((\mathbf{\Lambda}_1'\mathbf{\Psi}_1^{-2}\mathbf{\Lambda}_1)^{-1}\mathbf{\Lambda}_1'\mathbf{\Psi}_1^{-2}\mathbf{\Sigma}_1\mathbf{\Psi}_1^{-2}\mathbf{\Lambda}_1(\mathbf{\Lambda}_1'\mathbf{\Psi}_1^{-2}\mathbf{\Lambda}_1)^{-1})^{-1/2}\end{aligned} \tag{11}$$

It follows from $\operatorname{diag}(\mathbf{\Phi}_1) = \mathbf{I}$ that Equation 11 can be transformed into

$$\mathbf{R}_{ttb} = \operatorname{diag}((\mathbf{\Lambda}_1'\mathbf{\Psi}_1^{-2}\mathbf{\Lambda}_1)^{-1}\mathbf{\Lambda}_1'\mathbf{\Psi}_1^{-2}\mathbf{\Sigma}_1\mathbf{\Psi}_1^{-2}\mathbf{\Lambda}_1(\mathbf{\Lambda}_1'\mathbf{\Psi}_1^{-2}\mathbf{\Lambda}_1)^{-1})^{-1}. \tag{12}$$

Entering $\mathbf{\Lambda}_1\mathbf{\Phi}_1\mathbf{\Lambda}_1' + \mathbf{\Psi}_1^2$ for $\mathbf{\Sigma}_1$ into Equation 12 yields

$$\mathbf{R}_{ttb} = \operatorname{diag}((\mathbf{\Lambda}_1'\mathbf{\Psi}_1^{-2}\mathbf{\Lambda}_1)^{-1}\mathbf{\Lambda}_1'\mathbf{\Psi}_1^{-2}(\mathbf{\Lambda}_1\mathbf{\Phi}_1\mathbf{\Lambda}_1' + \mathbf{\Psi}_1^2)\mathbf{\Psi}_1^{-2}\mathbf{\Lambda}_1(\mathbf{\Lambda}_1'\mathbf{\Psi}_1^{-2}\mathbf{\Lambda}_1)^{-1})^{-1}, \tag{13}$$

and, after some transformation,

$$\mathbf{R}_{ttb} = \operatorname{diag}((\mathbf{\Lambda}_1'\mathbf{\Psi}_1^{-2}\mathbf{\Lambda}_1)^{-1} + \mathbf{\Phi}_1)^{-1}. \tag{14}$$

Entering $\mathbf{B}_m = \mathbf{\Psi}^{-2}\mathbf{\Lambda}\mathbf{N}(\mathbf{N}'\mathbf{\Lambda}'\mathbf{\Psi}^{-2}\mathbf{\Sigma}\mathbf{\Psi}^{-2}\mathbf{\Lambda}\mathbf{N})^{-1/2}$ when $\mathbf{N}$ is a $q \times q$ matrix with $\mathbf{N}\mathbf{N}' = \mathbf{\Phi}$ for McDonald's correlation preserving factor score predictor into Equation 6 and introducing subscripts and assuming $\mathbf{\Lambda}_1 = \mathbf{\Lambda}_2, \mathbf{\Phi}_1 = \mathbf{\Phi}_2, \mathbf{\Psi}_1 = \mathbf{\Psi}_2, \mathbf{\Sigma}_1 = \mathbf{\Sigma}_2, \mathbf{f}_1\mathbf{e}_2 = \mathbf{0}, \mathbf{f}_2\mathbf{e}_1 = \mathbf{0}$, and $\mathbf{e}_1\mathbf{e}_2' = \mathbf{0}$ yields

$$\begin{aligned}\mathbf{R}_{ttm} &= \operatorname{Cor}(\hat{\mathbf{f}}_{1m}, \hat{\mathbf{f}}_{2m}) \\ &= \operatorname{diag}((\mathbf{N}_1'\mathbf{\Lambda}_1'\mathbf{\Psi}_1^{-2}\mathbf{\Sigma}_1\mathbf{\Psi}_1^{-2}\mathbf{\Lambda}_1\mathbf{N}_1)^{-1/2}\mathbf{N}_1'\mathbf{\Lambda}_1'\mathbf{\Psi}_1^{-2}\mathbf{\Lambda}_1\mathbf{\Phi}_1\mathbf{\Lambda}_1'\mathbf{\Psi}_1^{-2}\mathbf{\Lambda}_1\mathbf{N}_1(\mathbf{N}_1'\mathbf{\Lambda}_1'\mathbf{\Psi}_1^{-2}\mathbf{\Sigma}_1\mathbf{\Psi}_1^{-2}\mathbf{\Lambda}_1\mathbf{N}_1)^{-1/2}).\end{aligned} \tag{15}$$

Thus, as for the Kuder-Richardson formula, only the parameters of the empirical items are necessary in order to calculate the reliabilities, when the hypothetical item set is equivalent. The equivalence of the items implies that the parameters of the factor model will be identical for the two item sets.

**Comparing reliability estimates for different factor score predictors**

It should be noted that the formula for the reliability estimated of factor score predictors are based on the condition of equal factor models and, especially, on $\mathbf{f}_1 = \mathbf{f}_2$, $\mathbf{f}_1\mathbf{e}_2' = \mathbf{0}$, $\mathbf{f}_2\mathbf{e}_1' = \mathbf{0}$, and $\mathbf{e}_1\mathbf{e}_2' = \mathbf{0}$. This means that all true variance and all reliability comes from the amount of variance that is due to $\mathbf{f}_1$. Thus, the factor score predictor with the highest correlation with $\mathbf{f}_1$ should have the highest reliability. The regression score predictor has the highest correlation with the factor [16], so that $\mathrm{Cor}(\hat{\mathbf{f}}_{1r},\mathbf{f}_1) \geq \mathrm{Cor}(\hat{\mathbf{f}}_{1b},\mathbf{f}_1)$ implies $\mathbf{R}_{ttr} \geq \mathbf{R}_{ttb}$, and $\mathrm{Cor}(\hat{\mathbf{f}}_{1r},\mathbf{f}_1) \geq \mathrm{Cor}(\hat{\mathbf{f}}_{1m},\mathbf{f}_1)$ implies $\mathbf{R}_{ttr} \geq \mathbf{R}_{ttm}$. Although the regression score predictor has the same or a larger reliability than the other two factor score predictors, the conditions for having an equal reliability are also of interest.

Theorem 1 shows that the reliabilities of the regression factor score predictor and the Bartlett factor score predictor are equal when the condition $\mathbf{\Lambda}_1'\mathbf{\Sigma}_1^{-1}\mathbf{\Lambda}_1 = \mathrm{diag}(\mathbf{\Lambda}_1'\mathbf{\Sigma}_1^{-1}\mathbf{\Lambda}_1)$ holds for orthogonal factor models ($\mathbf{\Phi}_1 = \mathbf{I}$). The conditions $\mathbf{\Lambda}_1'\mathbf{\Sigma}_1^{-1}\mathbf{\Lambda}_1 = \mathrm{diag}(\mathbf{\Lambda}_1'\mathbf{\Sigma}_1^{-1}\mathbf{\Lambda}_1)$ and $\mathbf{\Phi}_1 = \mathbf{I}$ hold for one-factor models, since $q = 1$ implies $\mathbf{\Phi}_1 = 1$ and that there is only one resulting number for $\mathbf{\Lambda}_1'\mathbf{\Sigma}_1^{-1}\mathbf{\Lambda}_1$. Moreover, the conditions $\mathbf{\Lambda}_1'\mathbf{\Sigma}_1^{-1}\mathbf{\Lambda}_1 = \mathrm{diag}(\mathbf{\Lambda}_1'\mathbf{\Sigma}_1^{-1}\mathbf{\Lambda}_1)$ and $\mathbf{\Phi}_1 = \mathbf{I}$ hold for orthogonal factor models when there is only one non-zero factor loading of each variable (perfect simple structure).

**Theorem 1.** *If* $\mathbf{\Lambda}_1 = \mathbf{\Lambda}_2, \mathbf{\Phi}_1 = \mathbf{\Phi}_2 = \mathbf{I}, \mathbf{\Psi}_1 = \mathbf{\Psi}_2, \mathbf{\Sigma}_1 = \mathbf{\Sigma}_2$, *and* $\mathbf{\Lambda}_1'\mathbf{\Sigma}_1^{-1}\mathbf{\Lambda}_1 = \mathrm{diag}(\mathbf{\Lambda}_1'\mathbf{\Sigma}_1^{-1}\mathbf{\Lambda}_1)$ *then* $\mathbf{R}_{ttr} = \mathbf{R}_{ttb}$.

*Proof.* From Jöreskog [17] (Equation 10) we get

$$\mathbf{\Sigma}_1^{-1}\mathbf{\Lambda}_1 = \mathbf{\Psi}_1^{-2}\mathbf{\Lambda}_1(\mathbf{I} + \mathbf{\Phi}_1\mathbf{\Lambda}_1'\mathbf{\Psi}_1^{-2}\mathbf{\Lambda}_1)^{-1}. \tag{16}$$

Premultiplication with $\mathbf{\Lambda}_1'$ and some transformation yields $\mathbf{\Lambda}_1'\mathbf{\Sigma}_1^{-1}\mathbf{\Lambda}_1 = ((\mathbf{\Lambda}_1'\mathbf{\Psi}_1^{-2}\mathbf{\Lambda}_1)^{-1} + \mathbf{\Phi}_1)^{-1}$ which is entered into Equation 9. This yields

$$\begin{aligned}\mathbf{R}_{ttr} = &\,\mathrm{diag}(\mathbf{\Phi}_1((\mathbf{\Lambda}_1'\mathbf{\Psi}_1^{-2}\mathbf{\Lambda}_1)^{-1} + \mathbf{\Phi}_1)^{-1}\mathbf{\Phi}_1)^{-1/2}\mathrm{diag}(\mathbf{\Phi}_1((\mathbf{\Lambda}_1'\mathbf{\Psi}_1^{-2}\mathbf{\Lambda}_1)^{-1} + \mathbf{\Phi}_1)^{-1}\\&\mathbf{\Phi}_1((\mathbf{\Lambda}_1'\mathbf{\Psi}_1^{-2}\mathbf{\Lambda}_1)^{-1} + \mathbf{\Phi}_1)^{-1}\mathbf{\Phi}_1)\mathrm{diag}(\mathbf{\Phi}_1((\mathbf{\Lambda}_1'\mathbf{\Psi}_1^{-2}\mathbf{\Lambda}_1)^{-1} + \mathbf{\Phi}_1)^{-1}\mathbf{\Phi}_1)^{-1/2}.\end{aligned} \tag{17}$$

According to the conditions of Theorem 1 Equation 17 can be transformed into

$$\mathbf{R}_{ttr} = \mathrm{diag}(((\mathbf{\Lambda}_1'\mathbf{\Psi}_1^{-2}\mathbf{\Lambda}_1)^{-1} + \mathbf{I})^{-1})^{-1/2}\mathrm{diag}(((\mathbf{\Lambda}_1'\mathbf{\Psi}_1^{-2}\mathbf{\Lambda}_1)^{-1} + \mathbf{I})^{-2})\mathrm{diag}(((\mathbf{\Lambda}_1'\mathbf{\Psi}_1^{-2}\mathbf{\Lambda}_1)^{-1} + \mathbf{I})^{-1})^{-1/2}. \tag{18}$$

Since $\mathbf{\Lambda}_1'\mathbf{\Sigma}_1^{-1}\mathbf{\Lambda}_1 = \mathrm{diag}(\mathbf{\Lambda}_1'\mathbf{\Sigma}_1^{-1}\mathbf{\Lambda}_1)$ and $\mathbf{\Phi}_1 = \mathbf{I}$ implies $((\mathbf{\Lambda}_1'\mathbf{\Psi}_1^{-2}\mathbf{\Lambda}_1)^{-1} + \mathbf{I})^{-1} = \mathrm{diag}((\mathbf{\Lambda}_1'\mathbf{\Psi}_1^{-2}\mathbf{\Lambda}_1)^{-1} + \mathbf{I})^{-1}$ Equation 18 can be transformed into

$$\mathbf{R}_{ttr} = \mathrm{diag}((\mathbf{\Lambda}_1'\mathbf{\Psi}_1^{-2}\mathbf{\Lambda}_1)^{-1} + \mathbf{I})^{-1}. \tag{19}$$

This completes the proof. ∎

Theorem 2 shows that the reliabilities of the regression factor score predictor and the McDonald factor score predictor are equal when the condition $\mathbf{\Lambda}_1'\mathbf{\Sigma}_1^{-1}\mathbf{\Lambda}_1 = \text{diag}(\mathbf{\Lambda}_1'\mathbf{\Sigma}_1^{-1}\mathbf{\Lambda}_1)$ holds for orthogonal factor models ($\mathbf{\Phi}_1 = \mathbf{I}$).

**Theorem 2.** *If* $\mathbf{\Lambda}_1 = \mathbf{\Lambda}_2, \mathbf{\Phi}_1 = \mathbf{\Phi}_2 = \mathbf{I}, \mathbf{\Psi}_1 = \mathbf{\Psi}_2, \mathbf{\Sigma}_1 = \mathbf{\Sigma}_2$, *and* $\mathbf{\Lambda}_1'\mathbf{\Sigma}_1^{-1}\mathbf{\Lambda}_1 = \text{diag}(\mathbf{\Lambda}_1'\mathbf{\Sigma}_1^{-1}\mathbf{\Lambda}_1)$ *then*
$\mathbf{R}_{ttr} = \mathbf{R}_{ttm}$.

*Proof.* For $\mathbf{\Phi}_1 = \mathbf{\Phi}_2 = \mathbf{I}$ Equation 15 can be written as

$$\mathbf{R}_{ttm} = \text{diag}\,((\mathbf{\Lambda}_1'\mathbf{\Psi}_1^{-2}\mathbf{\Sigma}_1\mathbf{\Psi}_1^{-2}\mathbf{\Lambda}_1)^{-1/2}\mathbf{\Lambda}_1'\mathbf{\Psi}_1^{-2}\mathbf{\Lambda}_1\mathbf{\Lambda}_1'\mathbf{\Psi}_1^{-2}\mathbf{\Lambda}_1(\mathbf{\Lambda}_1'\mathbf{\Psi}_1^{-2}\mathbf{\Sigma}_1\mathbf{\Psi}_1^{-2}\mathbf{\Lambda}_1)^{-1/2}). \tag{20}$$

Entering $\mathbf{\Lambda}_1\mathbf{\Lambda}_1' + \mathbf{\Psi}_1^2$ for $\mathbf{\Sigma}_1$ into Equation 20 and some transformation yields

$$\begin{aligned}\mathbf{R}_{ttm} &= \text{diag}\,((\mathbf{\Lambda}_1'\mathbf{\Psi}_1^{-2}\mathbf{\Lambda}_1\mathbf{\Lambda}_1'\mathbf{\Psi}_1^{-2}\mathbf{\Lambda}_1(\mathbf{\Lambda}_1'\mathbf{\Psi}_1^{-2}\mathbf{\Lambda}_1)^{-2} + \mathbf{\Lambda}_1'\mathbf{\Psi}_1^{-2}\mathbf{\Lambda}_1(\mathbf{\Lambda}_1'\mathbf{\Psi}_1^{-2}\mathbf{\Lambda}_1)^{-2})^{-1})\\
&= \text{diag}\,(((\mathbf{\Lambda}_1'\mathbf{\Psi}_1^{-2}\mathbf{\Lambda}_1)^{-1} + \mathbf{I})^{-1})\\
&= \text{diag}\,((\mathbf{\Lambda}_1'\mathbf{\Psi}_1^{-2}\mathbf{\Lambda}_1)^{-1} + \mathbf{I})^{-1}.\end{aligned} \tag{21}$$

This completes the proof. ∎

Thus, the three factor score predictors considered here have the same reliability for $q = 1$ and for orthogonal models with $q > 1$ and only one non-zero factor loading of each variable (perfect simple structure). However, these considerations do not allow for a quantification of the relative differences of the reliabilities of the factor score predictors. Therefore, simulation studies were performed in order to give an account of the reliabilities of the three factor score predictors under different conditions. First, a simulation study was performed at the level of the population for item sets for which the factor model holds in the population.

**Simulation Study 1.** The first short simulation study describes the effects of different population parameters on the reliability estimates. The simulation study was performed with IBM SPSS Version 22 and gives an account of the reliability estimates for the three factor score predictors for $q = 6$, depending on the number of main loadings per factor $p/q$ (5, 10), the size of main loadings $l$ (.40, .50, .60, .70, .80), the size of secondary loadings $sl$ (.00, .10), and the size of the factor inter-correlations $r$ (.00, .30). This results in (2 levels of $p/q$ × 5 levels of $l$ × 2 levels of $sl$ × 2 levels of $r$) 40 population models, for which population correlation matrices of observed variables were generated according to Equation 2. The models with $p/q = 5$ were based on 30 observed variables and the models with $p/q = 10$ were based on 60 observed variables.

The reliability estimates for the factor score predictors were computed from the population parameters of the factor model ($\mathbf{\Lambda}, \mathbf{\Phi}, \mathbf{\Psi}$) and the corresponding item covariances ($\mathbf{\Sigma}$) by means of Equations 9, 14, and 15. The results are summarized in Figure 1. No pronounced reliability differences occurred when the secondary loadings ($sl$) were zero, especially, when only reliabilities greater than .70 are considered. For $sl = .10$ and factor inter-correlations of .30, the regression score predictor had a notably larger reliability than Bartlett's factor score predictor and McDonald's factor score predictor. The differences between the reliability estimates for the Bartlett's factor score predictor and McDonald's factor score predictor were very small.

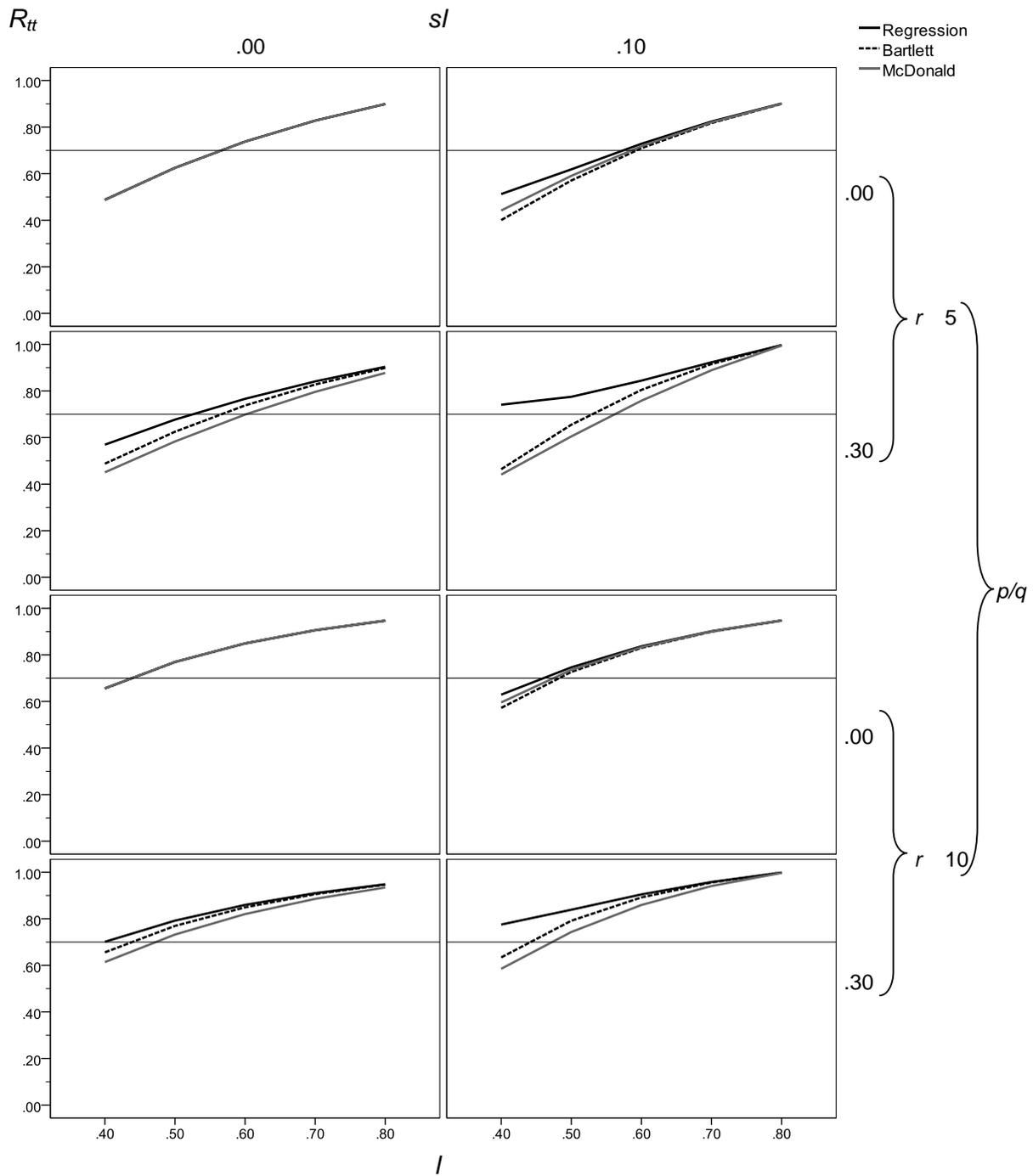

Fig. 1. Reliability estimates for the regression factor score predictor, Bartlett's factor score predictor, and McDonalds' factor score predictor for population models with $q = 6$. The horizontal line marks a reliability of .70 ($R_{tt}$ = Reliability estimate, $l$ = salient loadings, $sl$ = secondary loadings, $r$ = factor inter-correlations).

**Simulation Study 2.** The next simulation is based on samples that are drawn from populations with the same model parameters as in the previous simulation. The simulation study was again performed with IBM SPSS Version 22. For each of the 40 population models of the previous simulation study 1,000 samples with $n = 500$ cases and 1,000 samples with $n = 1,000$ cases were drawn. Random numbers for the samples of factor scores were generated by means of the SPSS

Mersenne Twister random number generator. The corresponding samples of observed variables were generated from the common and unique factor scores by means of Equation 2. Maximum-likelihood factor analysis with subsequent Varimax-rotation for orthogonal population factor models and with Promax-rotation (kappa=4) for correlated factor models was performed in each sample of observed variables and the corresponding factor score reliabilities were computed from Equations 9, 14, and 15. The results can be found in Figure 2.

The results of the simulation study for the samples are essentially the same as the results for the population parameters with the highest reliability of the regression factor score predictor. The main difference to the results of the simulation study for the population is that the Bartlett factor score predictor is substantially more reliable than the McDonald factor score predictor when the factor inter-correlations are substantial and when there are substantial secondary loadings.

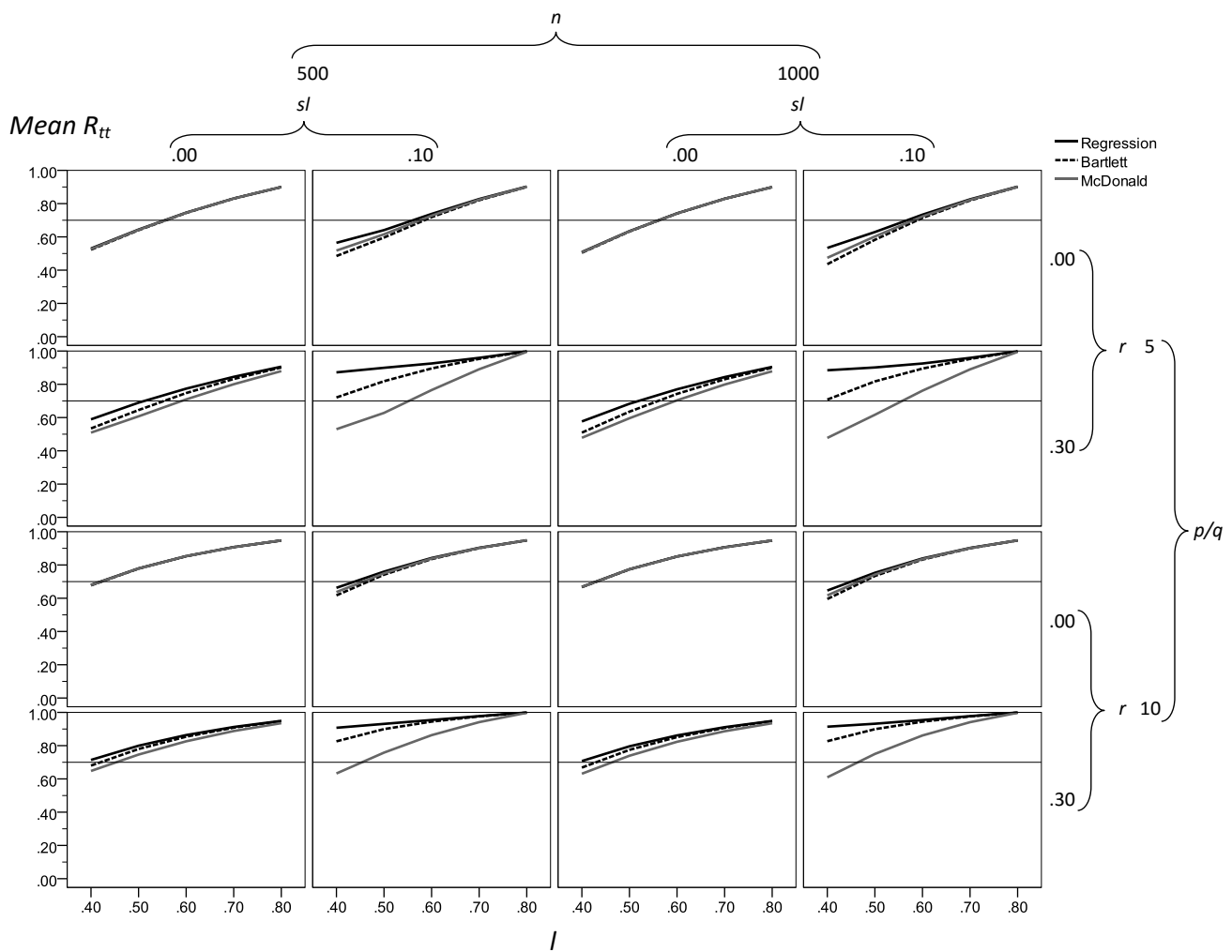

Fig. 2. Reliability estimates for the regression factor score predictor, Bartlett's factor score predictor, and McDonalds' factor score predictor for samples based on population models with $q = 6$. The horizontal line marks a reliability of .70 ($R_{tt}$ = Reliability estimate, $l$ = salient loadings, $sl$ = secondary loadings, $r$ = factor inter-correlations).

**Simulation Study 3.** The third simulation study was again based on the population parameters of the first and second simulation study. The only difference is that the simulation study was based on imperfect models, thus, on population models that do not fit exactly to the population covariance matrix [12,13]. Imperfect models were generated as proposed by MacCallum and Tucker [13]. The population correlation matrices were generated from the loadings of the major factors corresponding to the factors in the simulation studies 1 and 2 as well as from the loadings of 100 'minor factors' and from the corresponding uniquenesses. Minor factors have very small nonzero population loadings and represent the 'many minor influences', which are thought to affect the values of the observed scores in the real world. Again, maximum-likelihood factor analysis with subsequent Varimax-rotation for orthogonal population factor models and with Promax-rotation (kappa=4) for correlated factor models was performed in each sample of observed variables and the corresponding factor score reliabilities were computed. The results for the imperfect models were extremely similar to those presented in simulation study 2, so that an additional figure was not necessary. Thus, imperfect models did not affect the reliability estimates substantially.

**Reliability of the regression score predictor and the coefficient of determinacy**

In the following, the reliability estimate for the regression score predictor is compared with the determinacy coefficient [1] in order to give an account of the relation between reliability and validity. The covariances of the regression factor score predictor with the corresponding common factor are the diagonal elements of

$$\mathrm{diag}(\varepsilon[\mathbf{f_r f'}]) = \mathrm{diag}(\varepsilon[\mathbf{\Phi \Lambda' \Sigma^{-1} x f'}]) = \mathrm{diag}(\mathbf{\Phi \Lambda' \Sigma^{-1} \Lambda \Phi}). \tag{22}$$

The standard deviation of the factor is one and the standard deviation of the regression factor score predictor is $\mathrm{diag}(\mathbf{\Phi \Lambda' \Sigma^{-1} \Lambda \Phi})^{-1/2}$. Accordingly, the factor score determinacy, i.e., the correlation of the regression score predictor with the corresponding common factors is

$$\mathrm{diag}(\mathrm{cor}[\mathbf{f_r}, \mathbf{f}]) = \mathrm{diag}(\mathbf{\Phi \Lambda' \Sigma^{-1} \Lambda \Phi}) \, \mathrm{diag}(\mathbf{\Phi \Lambda' \Sigma^{-1} \Lambda \Phi})^{-1/2} = \mathrm{diag}(\mathbf{\Phi \Lambda' \Sigma^{-1} \Lambda \Phi})^{1/2}. \tag{23}$$

When the common variance of the factor and the regression factor score predictor is computed for the empirical factor models considered above, this yields

$$\mathrm{diag}\left(\mathrm{cor}[\mathbf{f_r}, \mathbf{f}]\right)^2 = \mathrm{diag}(\mathbf{\Phi_1 \Lambda_1' \Sigma_1^{-1} \Lambda_1 \Phi_1}). \tag{24}$$

For orthogonal factor models with $\mathbf{\Phi_1} = \mathbf{I}$ and $\mathbf{\Lambda_1' \Sigma_1^{-1} \Lambda_1} = \mathrm{diag}(\mathbf{\Lambda_1' \Sigma_1^{-1} \Lambda_1})$ Equation 9 can be transformed into

$$\begin{aligned}\mathbf{R_{ttr}} &= \mathrm{diag}(\mathbf{\Lambda_1' \Sigma_1^{-1} \Lambda_1})^{-1/2} \mathrm{diag}(\mathbf{\Lambda_1' \Sigma_1^{-1} \Lambda_1 \Lambda_1' \Sigma_1^{-1} \Lambda_1}) \mathrm{diag}(\mathbf{\Lambda_1' \Sigma_1^{-1} \Lambda_1})^{-1/2} \\ &= \mathrm{diag}(\mathbf{\Lambda_1' \Sigma_1^{-1} \Lambda}) = \mathrm{diag}\left(\mathrm{cor}[\mathbf{f_r}, \mathbf{f}]\right)^2.\end{aligned} \tag{25}$$

Thus, for orthogonal factor models with only one loading of each variable on one factor, the reliability estimate of the regression score predictor corresponds to the coefficient of determinacy. Since it has been shown that the reliability estimates of the regression score predictor, Bartlett's factor score predictor, and McDonald's factor score predictor are equal under these conditions, it follow that the abovementioned reliability estimates of the factor score predictors are equal to the determinacy coefficient for $\mathbf{\Phi}_1 = \mathbf{I}$ and $\mathbf{\Lambda}_1'\mathbf{\Sigma}_1^{-1}\mathbf{\Lambda}_1 = \text{diag}(\mathbf{\Lambda}_1'\mathbf{\Sigma}_1^{-1}\mathbf{\Lambda}_1)$.

Theorem 3 describes the relation between the reliability estimate of the regression factor score predictor and factor score determinacy for orthogonal factor models that are identical across measurement occasions when $\mathbf{\Lambda}_1'\mathbf{\Sigma}_1^{-1}\mathbf{\Lambda}_1 \neq \text{diag}(\mathbf{\Lambda}_1'\mathbf{\Sigma}_1^{-1}\mathbf{\Lambda}_1)$.

**Theorem 3.** *If* $\mathbf{\Lambda}_1 = \mathbf{\Lambda}_2, \mathbf{\Phi}_1 = \mathbf{\Phi}_2 = \mathbf{I}, \mathbf{\Psi}_1 = \mathbf{\Psi}_2, \mathbf{\Sigma}_1 = \mathbf{\Sigma}_2$, *and* $\mathbf{\Lambda}_1'\mathbf{\Sigma}_1^{-1}\mathbf{\Lambda}_1 \neq \text{diag}(\mathbf{\Lambda}_1'\mathbf{\Sigma}_1^{-1}\mathbf{\Lambda}_1)$ *then* $\mathbf{R}_{ttr} \geq \text{diag}\left(\text{cor}[\mathbf{f}_r, \mathbf{f}]\right)^2$.

*Proof.* For simplification we introduce $\text{diag}(\mathbf{\Lambda}_1'\mathbf{\Sigma}_1^{-1}\mathbf{\Lambda}_1) = \mathbf{D}$ and $\mathbf{\Lambda}_1'\mathbf{\Sigma}_1^{-1}\mathbf{\Lambda}_1 - \mathbf{D} = \mathbf{H}$.

Accordingly, Equation 9 can be written as

$$\mathbf{R}_{ttr} = \mathbf{D}^{-1/2}\text{diag}(\mathbf{HH} + \mathbf{HD} + \mathbf{DH} + \mathbf{DD})\,\mathbf{D}^{-1/2}. \tag{26}$$

Since $\mathbf{H}$ has a zero-diagonal, pre- and post-multiplication of $\mathbf{H}$ with the diagonal matrix $\mathbf{D}$ does not alter the diagonal elements, so that the diagonal elements in $\mathbf{HD}$ and $\mathbf{DH}$ are zero. Therefore, Equation 26 can be written as

$$\mathbf{R}_{ttr} = \mathbf{D}^{-1/2}\text{diag}(\mathbf{HH} + \mathbf{DD})\,\mathbf{D}^{-1/2}. \tag{27}$$

Since these diagonal elements are squared elements, it follows that

$$\text{diag}(\mathbf{HH}) \geq 0 \text{ and } \text{diag}(\mathbf{DD}) \geq 0. \tag{28}$$

For orthogonal models Equation 24 can be written as

$$\text{diag}\left(\text{cor}[\mathbf{f}_r, \mathbf{f}]\right)^2 = \mathbf{D} = \mathbf{D}^{-1/2}\mathbf{DD}\,\mathbf{D}^{-1/2}. \tag{29}$$

It follows from $\text{diag}(\mathbf{HH}) \geq 0$ that $\mathbf{R}_{ttr} \geq \text{diag}\left(\text{cor}[\mathbf{f}_r, \mathbf{f}]\right)^2$.

This completes the proof. ∎

To summarize, the determinacy coefficient, i.e., $\text{diag}\left(\text{cor}[\mathbf{f}_r, \mathbf{f}]\right)^2$, corresponds to the reliability of the regression score predictor for orthogonal factor models that are identical across measurement occasions (when $\mathbf{\Lambda}_1'\mathbf{\Sigma}_1^{-1}\mathbf{\Lambda}_1 = \text{diag}(\mathbf{\Lambda}_1'\mathbf{\Sigma}_1^{-1}\mathbf{\Lambda}_1)$) and $\text{diag}\left(\text{cor}[\mathbf{f}_r, \mathbf{f}]\right)^2$ is a lower-bound estimate of the reliability of the regression score predictor when $\mathbf{\Lambda}_1'\mathbf{\Sigma}_1^{-1}\mathbf{\Lambda}_1 \neq \text{diag}(\mathbf{\Lambda}_1'\mathbf{\Sigma}_1^{-1}\mathbf{\Lambda}_1)$, which can occur when there are non-zero secondary loadings.

**Discussion**

Reliability estimates for Thurstone's regression factor score predictor, Bartlett's factor score predictor, and McDonald's factor score predictor were proposed. As in Kuder-Richardson's formula, the reliability estimates are based on a hypothetical set of equivalent items. The reliability estimates were, moreover, based on the assumption that the true variance of the items is only based on the common factors and that the error or unique variances of the items due not contribute to the reliability of the factor score predictors. Other assumptions might be possible, e.g. for hierarchical factor models, when the unique variance of a second order factor analysis already represents some amount of true score variance. However, this is not the standard case and the aim of the present study was to propose reliability estimates for the common case. It was shown that the reliability estimates are equal for the three factor score predictors when they are based on a one-factor model or when there are orthogonal factors with only one non-zero loadings of the items on a factor.

The reliability estimates of the three factor score predictors were compared by means of a simulation study for the population and by means of a simulation study for samples drawn from a population in which the factor model holds as well as for samples drawn from a population in which the factor model does not hold. It was found in the population based simulation study that the reliability estimates were largest for the regression factor score predictor and that the differences between the reliability estimates for Bartlett's factor score predictor and McDonald's factor score predictor were small. Especially, for models with correlated factors and substantial secondary loadings, the regression factor score predictor had substantially larger reliability estimates. In contrast, for orthogonal factors and when only substantial reliabilities (>.70) were considered, the differences between the reliability estimates for all three factor score predictors were small. The results of the simulation studies for the samples were very similar to the results for the population based simulation study. The regression factor score predictor was most reliable across all conditions. At best, the Bartlett and McDonald factor score predictor were as reliable as was the regression factor score predictor. The only relevant difference between the sample based simulation studies and the population based simulation study was that the Bartlett factor score predictor was substantially more reliable than the McDonald factor score predictor when there were substantial factor inter-correlations and non-zero secondary loadings in the sample based simulation study. Thus, computing McDonald's factor score predictor may result in larger losses of reliability than computing Bartlett's factor score predictor. The effect of using imperfect factor models for the simulation study did not affect the results.

Overall, the results of the simulation studies indicate that whenever Bartlett's or McDonald's factor score predictor are to be computed, the resulting reliability estimates should be compared with the reliability of the regression factor score predictor. This is necessary in order to investigate whether a substantial amount of reliability is lost by computing Bartlett's or McDonald's factor score predictor instead of the regression factor score predictor. An R-script (Appendix A) as well as an SPSS-script (Appendix B) was presented that allows for the respective calculations of the reliability estimates from the loading pattern and factor inter-correlations.

Finally, it was shown that the reliability estimates for the regression factor score predictor are equal to the determinacy coefficient for the one-factor model or when there are orthogonal factors with only one non-zero loadings of the items on a factor. For orthogonal factor models with more than one non-zero loading of the items on a factor the determinacy coefficient is a lower-bound estimate of the reliability of the regression factor score predictor. This result was not unexpected since the determinacy coefficient is based on the correlation of the regression factor score predictor with the factor.

**Appendix A**

**R-script for reliabilities of factor score predictors**

```
##' This function computes and returns reliability estimates for three commonly used
##' Factor Score Predictors in Factor Analyses.
##'
##' Explanations of the algebraic formulas are presented in the manuscript.
##'
##' @title Function for calculating reliability estimates for factor score predictors
##' @param Lambda a \code{matrix} containing the loadings of items on the factors
##' @param Phi a \code{matrix} containing the factor intercorrelations
##' @param Predictors a \code{vector} to select the predictors for which the reliability
##'     estimates should be calculated. Available values: \code{Regression}, \code{Bartlett},
##'     \code{McDonald}
##' @return Returns a two-dimensional list containing the reliability estimates for each
##'     factor. Depending on the \code{Predictors} parameter, the list contains the values
##'         only for the selected Predictors.
##' @export
##' @author André Beauducel (\email{beauducel@uni-bonn.de})
##' @author Christopher Harms (\email{christopher.harms@uni-bonn.de})
##' @author Norbert Hilger (\email{nhilger@uni-bonn.de})
##'

factor.score.reliability <- function(Lambda, Phi, Predictors=c("Regression", "Bartlett", "McDonald")) {
    # Helper functions for frequently used matrix operations
    Mdiag <- function(x) return(diag(diag(x)))
    inv <- function(x) return(solve(x))

    # If a 'loadings' class is provided for lambda, we can easily convert it
    if (is(Lambda, "loadings"))
        Lambda <- Lambda[,]

    # Perform several validity checks of the provided arguments
    if (any(missing(Lambda), missing(Phi), is.null(Lambda), is.null(Phi)))
        stop("Missing argument(s).")
    if (any(nrow(Phi) == 0, nrow(Lambda) == 0, ncol(Phi) == 0, ncol(Lambda) == 0))
        stop("Some diemension(s) of Phi or Lambda seem to be empty.")
    if (nrow(Phi) != ncol(Phi))
        stop("Phi has to be a q x q matrix.")
    if (ncol(Lambda) != nrow(Phi))
        stop("Phi and Lambda have a different count of factors.")
    if (any(round(min(Phi)) < 0, round(max(Phi)) > 1))
        stop("Phi contains invalid values (outside [0; 1]).")
    Predictors.Allowed <- c("Regression", "Bartlett", "McDonald")
    if (is.null(Predictors)) {
        message("No 'Predictors' defined, use 'Regression' as default.")
        Predictors <- c("Regression")
    }
    Predictors <- match.arg(Predictors, Predictors.Allowed, several.ok = TRUE)
```

```
        # Regenerate covariance matrix from factor loadings matrix
        Sigma <- (Lambda %*% Phi %*% t(Lambda))
        Sigma <- Sigma - Mdiag(Sigma) + diag(nrow(Lambda))

        # Calculate uniqueness/error of items
        Psi <- Mdiag(Sigma - Lambda %*% Phi %*% t(Lambda))^0.5
        if (round(min(diag(Psi))) < 0)
            stop("The diagonal of Psi contains negative values.")

    ret <- list()
if ("Regression" %in% Predictors) {
            # Reliability of Thurstone's Regression Factor Score Predictors
            # cf. Equation 9 in manuscript
Rtt.Regression <-
inv( Mdiag( Phi %*% t(Lambda) %*% inv(Sigma) %*% Lambda %*% Phi ) )^0.5 %*%
Mdiag( Phi %*% t(Lambda) %*% inv(Sigma) %*% Lambda %*% Phi %*% t(Lambda) %*%
inv(Sigma) %*% Lambda %*% Phi) %*%
inv( Mdiag( Phi %*% t(Lambda) %*% inv(Sigma) %*% Lambda %*% Phi ) )^0.5
            ret$Regression <- diag(Rtt.Regression)
        }
if ("Bartlett" %in% Predictors) {
            # Reliability of Bartlett's Factor Score Predictors
            # cf. Equation 14 in manuscript
Rtt.Bartlett <- inv( Mdiag( inv(t(Lambda) %*% inv(Psi)^2 %*% Lambda) + Phi ) )
            ret$Bartlett <- diag(Rtt.Bartlett)
        }
if ("McDonald" %in% Predictors) {
            # Reliability of McDonald's correlation preserving factor score predictors
            # cf. Equation 15 in manuscript
            Decomp <- svd(Phi)
            N <- Decomp$u %*% abs(diag(Decomp$d))^0.5
            sub.term <-
t(N) %*% t(Lambda) %*% inv(Psi)^2 %*% Sigma %*% inv(Psi)^2 %*% Lambda %*% N
            Decomp <- svd(sub.term)
            sub.term <- Decomp$u %*% (diag(Decomp$d)^0.5) %*% t(Decomp$u)
Rtt.McDonald <-
Mdiag( inv(sub.term) %*% t(N) %*% t(Lambda) %*% inv(Psi)^2 %*% Lambda %*% Phi %*%
t(Lambda) %*% inv(Psi)^2 %*% Lambda %*% N %*% inv(sub.term))
            ret$McDonald <- diag(Rtt.McDonald)
        }

# Return reliabilities as list, so it can be accessed via e.g. factor.score.reliability(L, P)$Regression
    return(ret)
}

## Example 1:
## Users may just enter their respective values for Loadings and InterCorr.
Loadings <- matrix(c(
    0.50,-0.10, 0.10,
    0.50, 0.10, 0.10,
```

```
       0.50, 0.10,-0.10,
      -0.10, 0.50, 0.15,
       0.15, 0.50, 0.10,
      -0.15, 0.50, 0.10,
       0.10, 0.10, 0.60,
       0.10,-0.10, 0.60,
       0.10, 0.10, 0.60
      ),
      nrow=9, ncol=3,
      byrow=TRUE)
InterCorr <- matrix(c(
      1.00, 0.30, 0.20,
      0.30, 1.00, 0.10,
      0.20, 0.10, 1.00
      ),
      nrow=3, ncol=3,
      byrow=TRUE)

reliabilities <- factor.score.reliability(Lambda = Loadings, Phi = InterCorr, Predictors = c("Regression", "Bartlett", "McDonald"))
lapply(reliabilities, round, 3)
```

**Appendix B**

**SPSS-script for reliabilities of factor score predictors**

```
* ' This function computes and returns reliability estimates for three commonly used
  ' Factor Score Predictors in Factor Analyses,
  '
  ' Explanations of the algebraic formulas are presented in the manuscript
  '
  ' André Beauducel (\email{beauducel@uni-bonn.de})
  ' Christopher Harms (\email{christopher.harms@uni-bonn.de})
  ' Norbert Hilger (\email{nhilger@uni-bonn.de})
/*.

MATRIX.

* Users may enter their respective numbers into the loading matrix:.
compute L={
 0.50,-0.10, 0.10;
 0.50, 0.10, 0.10;
 0.50, 0.10,-0.10;
-0.10, 0.50, 0.15;
 0.15, 0.50, 0.10;
-0.15, 0.50, 0.10;
 0.10, 0.10, 0.60;
 0.10,-0.10, 0.60;
 0.10, 0.10, 0.60
}.
print L/format=F5.2.

* Enter respective numbers into factor inter-correlations.
compute Phi={
 1.00, 0.30, 0.20;
 0.30, 1.00, 0.10;
 0.20, 0.10, 1.00
}.
print Phi/format=F5.2.

* Reproduce the observed covariances from the parameters of the factor model.
compute Sig=L*Phi*T(L).
compute Sig=Sig-Mdiag(diag(Sig))+ident(nrow(L),nrow(L)).

* Spezigität/Uniqueness/Error der Items berechnen.
compute Psi=Mdiag(diag(Sig-L*Phi*T(L)))&**0.5.

* Equation 9.
compute Rtt_r = INV( Mdiag(diag( Phi*T(L)*INV(Sig)*L*Phi )) )&**0.5 *
Mdiag(diag(Phi*T(L)*INV(Sig)*L*Phi*T(L)*INV(Sig)*L*Phi)) *
INV(Mdiag(diag(Phi*T(L)*INV(Sig)*L*Phi)))&**0.5 .
```

```
* Equation 14.
compute Rtt_b=INV( Mdiag(diag(INV(T(L)*INV(Psi)&**2*L) + Phi)) ).

* Equation 15.
CALL svd(phi, QQ, eig, QQQ).
compute N=QQ*abs(eig)&**0.5.
compute help=T(N)*T(L)*INV(Psi)&**2*Sig*INV(Psi)&**2*L*N.
CALL svd(help, QQ, eig, QQQ).
compute help12=QQ*((eig)&**0.5)*T(QQ).

compute Rtt_m=Mdiag(diag(
INV(help12)*T(N)*T(L)*INV(Psi)&**2*L*Phi*T(L)*INV(Psi)&**2*L*N*INV(help12)
)).

print/Title "Reliabilities for Regression factor score predictors:".
print {T(diag(rtt_r))}/Format=F6.3.
print/Title "Reliabilities for Bartlett factor score predictors:".
print {T(diag(rtt_b))}/Format=F6.3.
print/Title "Reliabilities for McDonald factor score predictors:".
print {T(diag(rtt_m))}/Format=F6.3.

END MATRIX.
```